\documentclass[twocolumn,prl,reprint,letterpaper]{revtex4-1}

\usepackage{graphicx}
\usepackage{dcolumn}
\usepackage{bm}
\usepackage{amsmath}
\usepackage{amssymb}
\usepackage{amsfonts}
\usepackage{color}
\usepackage{psfrag}
\usepackage{epstopdf}
\usepackage{natbib}
\usepackage{appendix}
\usepackage{lmodern}

\def\cite{\citep}

\newcommand{\A}{\mathsf{A}}
\newcommand{\T}{\mathsf{T}}
\newcommand{\G}{\mathsf{G}}
\newcommand{\C}{\mathsf{C}}

\begin{document}

\title{A simple biophysical model predicts more rapid accumulation of hybrid incompatibilities in small populations}


\author{Bhavin S. Khatri}

\email{bhavin.khatri@physics.org}

\affiliation{%
MRC National Institute for Medical Research\\
Mathematical Biology Division\\
The Ridgeway, London, NW7 1AA, U.K.
}%
\author{Richard A. Goldstein}%

\affiliation{%
Division of Infection and Immunity\\
University College London\\
London, WC1E 6BT, U.K.
}%

\begin{abstract}

Speciation is fundamental to the huge diversity of life on Earth. Evidence suggests reproductive isolation arises most commonly in allopatry with a higher speciation rate in small populations. Current theory does not address this dependence in the important weak mutation regime.  Here, we examine a biophysical model of speciation based on the binding of a protein transcription factor to a DNA binding site, and how their independent co-evolution, in a stabilizing landscape, of two allopatric lineages leads to incompatibilities. Our results give a new prediction for the monomorphic regime of evolution, consistent with data, that smaller populations should develop incompatibilities more quickly. This arises as: 1) smaller populations having a greater initial drift load, as there are more sequences that bind poorly than well, so fewer substitutions are needed to reach incompatible regions of phenotype space; 2) slower divergence when the population size is larger than the inverse of discrete differences in fitness. Further, we find longer sequences develop incompatibilities more quickly at small population sizes, but more slowly at large population sizes. The biophysical model thus represents a robust mechanism of rapid reproductive isolation for small populations and large sequences, that does not require peak-shifts or positive selection.

\end{abstract}

\maketitle

\section*{Introduction}

Speciation is of great importance in generating the observed diversity of life, yet it is still poorly understood, especially at the genetic level. Darwin \cite{Darwin1859_Hybridism}, despite the title of his magnum opus, struggled with the following problem, here phrased in a modern context: if hybrid inviability were due to heterozygote disadvantage at a single locus with alleles $a$ and $A$, it is difficult to see how two species could have evolved from a single homozygotic ancestor without going through the inviable heterozygotic state. A resolution of this paradox was to propose that non-linear interactions or epistasis between different loci can give rise to so-called Dobzhansky-Muller incompatibilities (DMI) \cite{Dobzhansky1936a,Muller1942,Bateson1909}; for example, two geographically isolated lineages evolving allopatrically from a common ancestor \textit{ab} can fix the allelic combinations \textit{aB} and \textit{Ab} respectively, yet the hybrid genotype $AB$ can be inviable. It has also been recognised that a different sort of hybrid incompatibility can arise in polygenic systems, where many loci code for an additive quantitative trait. Quadratic selection induces epistasis such that divergent populations, under the action of drift, maintain different underlying allelic combinations at the many loci \cite{Wright1935,Wright1935+} for the same optimal trait value, which when combined in hybrids can lead to incompatibilities \cite{Barton1989}. Field data \cite{CoyneOrrSpeciationBook2004,MayrBook1963, Seehausen2014} suggest that the dominant mechanism for the evolution of (postzygotic) reproductive isolation (RI) involves the accumulation of Dobzhansky-Muller incompatibilities in geographically isolated populations with no or little gene flow  \cite{Wu1983,Vigneault1986}. 

Despite many studies of the evolution of RI, very little attention has been paid to the role of population size; however, there is indirect evidence that smaller populations develop incompatibilities more quickly. In particular, observations of the large species diversity in small habitats \cite{Santos2012,MayrInBook1970,Glor2004}, such as cichlids in the East African Great Lakes \cite{Owen1990}, contrasted to the lower diversity of marine animals \cite{MayrInBook1970, Mayr1954,Rubinoff1971} and birds \cite{Fitzpatrick2004}, which have large ranges and population sizes, suggest that the rate at which RI develops increases with decreasing populations size. More directly, cichlids, whose effective population size is of order $100-10000$ \cite{vanOppen1997,Fiumera2000}, develop reproductive isolation on a timescale of $1-10$Myr after divergence \cite{Stelkens2010}, whilst domestic chickens (\textit{Gallus gallus}) can still hybridize with helmeted guineafowl (\textit{Numida meleagris}), even after roughly $55$Myr divergence \cite{Cooper1997a}, where estimates of the effective population size of domestic chickens range between $N_e\approx10^5$ to $10^6$ \cite{Sawai2010}. This trend is further supported \cite{CoyneOrrSpeciationBook2004} by inference of net diversification rates from phylogenetic trees \cite{Nee2001,Barraclough2001}.

Where does current theory stand in light of these observations?  There are a number of theoretical models of allopatric speciation based on the Dobzhansky-Muller mechanism, which consider independent lineages evolving neutrally or under varying selection pressures on each lineage \cite{Orr1995,Orr2001,OrrOrr1996,Nei1983,Gavrilets2003,Gavrilets1999, GavriletsBook}. Models which involve positive selection driving divergence are unlikely to be able to explain this dependence on population size, since larger populations respond more quickly, for a given selection pressure \cite{Gavrilets2003}. This leaves models of speciation where populations diverge neutrally or under similar stabilizing selection pressure; the models of Nei et. al \cite{Nei1983} and Gavrilets \cite{Gavrilets1999} tackle precisely this question in the strong mutation regime ($n\mu_0 N\geq 1$, where $n$ is the number of nucleotides or base-pairs for the loci of interest, $\mu_0$ the base-pair mutation rate and $N$ the population size). They find slower divergence in larger populations due to the lower mating success of members of the population who have diverged an amount comparable to the width of the fitness peak, resulting in a slower rate of developing RI. However, in neither of these models is there a dependence on population size in the weak mutation, nearly monomorphic regime, where $n\mu_0N\ll 1$. Models of hybrid incompatibility that rely on fitness epistasis on quantitative traits, also predict that smaller populations should develop reproductive isolation more quickly, as drift helps populations shift between stable equilibria more rapidly \cite{Barton1989}; but again by it's polygenic nature, we expect these results to be only relevant to the strong mutation regime. Johnson and Porter \cite{Johnson2000,Johnson2007a} examined the evolution of decreased hybrid fitness for simple models of gene regulation, under positive and stabilizing selection, in the clonal interference regime ($n\mu_0N\sim1$), but did not investigate the dependence on population size. More recently, they have extended their work with sequence based models \cite{Tulchinsky2014}, showing decreased hybrid fitness with decreasing population size, however, these results are again in the regime where the effect of mutations are not weak ($n\mu_0N\sim1$) and they did not investigate in detail the dynamics of the growth of DMIs. A model which could give rise to more rapid RI for small populations is based on founder events or peak-shifts, where small founder populations split and become isolated \cite{Lande1979,Lande1985a,Barton1984,Barton1987}; here genetic drift allows smaller populations to pass more easily through a fitness valley. A major problem with such models is that for isolation to occur on reasonable timescales the product of the fitness barrier and population size needs to be sufficiently small. However, this condition also means that gene flow is relatively unimpeded between peaks \cite{CoyneOrrSpeciationBook2004}, destroying the reproductive isolation the model seeks to establish. Finally, the work of Orr and co-workers, provided a framework to understand how incompatibilities might arise in allopatry through sequentially fixing mutations in the weak mutation regime ($n\mu_0 N\ll 1$) \cite{Orr1995,Orr2001}; they showed that the number of potential or untested incompatibilities ``snowballs'' like $\sim K^2$ for interactions between pairs of loci. However, the starting point of this model is the assumption of neutral, population-size independent, divergence between lineages with a fixed probability that each untested combination is incompatible, and so cannot address the question of the population size dependence. In summary, although the models of Gavrilets, Nei and Barton each predict a decreasing rate of developing RI with increasing population size when $n\mu_0 N\geq 1$, these models predict no dependence on population size, or are not applicable in the weak mutation, nearly monomorphic regime where $n\mu_0N\ll 1$. This is despite genetic studies which have shown that traits involved in species differences range from monogenic through to mildly polygenic \cite{Orr2001a}. For traits which are not very polygenic, ($n\mu_0 N\ll1$), we still lack an understanding of the effect of population size on the rate at which RI arises.

In this paper, we examine how incompatibilities arise in allopatry for an abstract, yet biophysically motivated model of binding between two macromolecules, a protein transcription factor (TF) binding to a specific DNA or TF binding site (TFBS). Our model is based on the ``two-state'' approximation \cite{Hippel1986,Gerland2002a}, which although not capturing the molecular interactions in atomistic detail, can represent many salient aspects which have been ignored in previous work on speciation theory. In particular, such a model allows us to include the effects of drift-selection balance, due to some phenotypes being coded by more sequences than others, and the corresponding effect of population size on the speciation dynamics in the weak mutation regime ($n\mu_0 N\ll1$). Recent work has shown that such mappings from genotype to phenotype give rise to a number of non-trivial effects \cite{Fontana2002,Force1999,Berg2004,Khatri2009,Mustonen2005,Goldstein2011}. Here, we find this simple genotype-phenotype map predicts an increasing rate of accumulating DMIs for decreasing population sizes in the weak mutation regime, the appropriate limit for monomorphically evolving traits, with a robust mechanism that does not require valley crossing by either of the divergent populations. This dependence on population size arises due to the fact that there are many more sequences that give weaker binding than good, so the common ancestor of smaller populations, which are dominated by genetic drift, are on average closer to the inviability boundary. 

Gene expression divergence has been shown to be a major factor in driving differences between species \cite{King1975,Wolf2010,Wray2007,Wittkopp2008}, providing indirect evidence of a role in speciation. In particular, compensatory changes at both $cis$ and $trans$ locations has been shown to be responsible for the misexpression of many genes in hybrids between \textit{D.melanogaster} and \textit{D.simulans} \cite{Landry2005}, as well as more direct evidence of speciation driven by the evolution of genes related to transcription factors in \textit{Drosophila} \cite{Ting1998,Brideau2006}. With the increasing use of genome level studies \cite{Seehausen2014} to study the process of speciation, there is a need for theory and modelling to bridge the gap between sequence level changes at co-evolving loci and phenotypic determinants of incompatibilities; the binding of transcription factors to DNA to control gene expression is arguably one of the most important co-evolving systems for organisms and so makes an ideal case study to examine the consequences to speciation of a simple biophysical model and a first mechanistic insight on the way DMIs develop.

The paper is organized as follows: We first introduce a biophysical model of a transcription factor binding to DNA and the population genetic model of their evolution. We then consider two populations evolving independently, and consider the viability of reproductive crosses between these populations.

\section*{Methods}

\subsection*{Quaternary Model of Transcription Factor-DNA Binding}

The two-state approximation \cite{Hippel1986,Gerland2002a} for transcription factor binding assumes that amino acid nucleotide interactions are either optimal or non-optimal and the contribution of each to the total binding energy is approximately additive. The rationale for this model is the underlying biophysics of protein-DNA interactions, in particular, the fact that an amino acid at a protein DNA interface will tend to have a preferred nucleotide with which to hydrogen bond, taking account the approximately fixed orientation of the amino acid as positioned by the rest of the protein. The other nucleotides tend to be non-optimal and not able to hydrogen bond \cite{Takeda1989}. Although each optimal interaction is marginally stabilizing ($-0.5$kcal/mol \cite{Hippel1986}), it is the non-optimal nucleotides that dominate the binding free energy, since they can neither hydrogen bond to an amino acid nor to water molecules. Although this suggests a large cost for each non-optimal interaction, in reality this is highly dependent on the particular protein and DNA sequence; the cost in free energy per amino acid nucleotide mismatch can range from 1-2 kcal/mol (2-3$k_BT$) \cite{Stormo1998,Takeda1989} to 4-5 kcal/mol (6-8$k_BT$) \cite{Lesser1990,Hippel1986,Baldwin2003}. This is likely explained by specific co-operative effects that include electrostatic, steric and solvent interactions \cite{Lesser1990,Baldwin2003} that change the energy scale of binding dependent on a particular protein-DNA binding context. In this paper, for simplicity, we assume that $\epsilon=3k_BT$.

As mentioned, for each amino acid there tends to a single nucleotide it prefers to hydrogen bond \cite{Takeda1989}. If we designate the category of amino acids by its preferred partnering base (e.g. an amino acid in group $\T$ would interact preferably with a thymine), and recognize that only changes of amino acid group affect the binding properties, we can use $\A$, $\T$, $\C$, and $\G$ to represent letters from the quaternary alphabet for both proteins and DNA sequences; for simplicity, this assumes that the amino acids are equally distributed amongst the four categories. In this way, the genome corresponding to this binding protein - binding location pair consists of two `genes' of length $\ell$ in the standard four letter alphabet of DNA.

For simplicity, we can then let the binding free energy be proportional to the number of amino acid-DNA mismatches, equal to the Hamming distance $r=d_H(\boldsymbol{g}^P,\boldsymbol{g}^D)$, where the function $d_H$ counts the number of positions where the two sequences $\boldsymbol{g^P}$ and $\boldsymbol{g^D}$ are not the same:

\begin{equation}\label{Eqn:BindingEnergy}
\Delta G=\varepsilon r,
\end{equation}
where $\varepsilon$ is the energy scale for a given transcription factor. This binding free energy corresponds to the specifically bound mode of attachment (which has both specific and non-specific contributions) and is in thermodynamic competition with the non-specifically bound mode of attachment, which is purely electrostatic. The free energy of binding in the electrostatic non-specific mode is, 

\begin{equation}\label{Eqn:NonSpecific_BindingEnergy}
\Delta G_{ns}=\ell\Delta\varepsilon_{ns}.
\end{equation}
where $\Delta\varepsilon_{ns}$ is the effective increase in free energy per nucleotide in the non-specific mode relative to the best binder. Thermodynamic studies of Lac repressor binding to DNA suggest that the difference in free energy between the best specific binding and the non-specific mode of binding is roughly $15k_BT$, so as $\ell=10$ for the Lac respressor, we find $\Delta\varepsilon_{ns}\approx 1.5k_BT$ \cite{Hippel1986,Revzin1977}. As the number of mismatches increases between a DNA binding site and protein sequence, the probability of the non-specific mode of attachment increases, which we assume leads to a decrease in functionality of the site; for simplicity, we model this below using truncation selection with a critical number of mismatches $r^*$.

\subsection*{TF-DNA binding evolution}

Relating the binding energy of a TF to its binding site to the fitness of an organism is in principle very complicated. In general, we would expect that in order for a TF to find its binding site it would need to minimise the number of mismatches and so typically we might expect that the fitness will increase with decreasing $r$. This is further supported by genome-wide studies of the distribution of binding energies for different TFs in \textit{E.coli} \cite{Mustonen2005} and yeast \cite{Mustonen2008,Haldane2014}, which show that there is a deviation of this distribution from the random/neutral expectation (Eqn.\ref{Eqn:Omega} below) for the best or lowest affinity binders. This deviation from the neutral distribution is related to selection for functional binding sites and has a character that suggests an effective (Malthusian) fitness landscape for binding energies, which is peaked with negative curvature. For simplicity, we therefore assume a quadratic log-fitness landscape, which is equivalent to a Gaussian fitness landscape (also referred to as a Wrightian fitness landscape or sometimes as a Darwinian fitness landscape \cite{HartlClarkBook2007}). To model competition between the specific and non-specific modes of attachment, we assume there is a critical number of mismatches $r^*$, where the probability of binding in each mode is equal; this happens when $\Delta G(r)=\Delta G_{ns}$, which from Eqn.\ref{Eqn:BindingEnergy} and Eqn.\ref{Eqn:NonSpecific_BindingEnergy} gives $r^*=\ell\Delta\varepsilon_{ns}/\Delta\varepsilon$. This simply says that that as binding sites increase in length, $\ell$, the stability of the best binder ($r=0$) relative to non-specific binding will increase in proportion to $\ell$ and hence a larger number of mismatches will be required before a binding site becomes non-functional. Specifically, for $\Delta\varepsilon=3k_BT$ and $\Delta\varepsilon_{ns}=1.5k_BT$ \cite{Hippel1986}, we get the relation $r^*=\ell/2$. In the case of short DNA recognition sites for Eco RI endonuclease cleaving DNA, where $\ell=5$, it was found that $r^*\approx3$ \cite{Lesser1990}, which agrees well with our approximate relation between $r^*$ and $\ell$. Thus we set the log-fitness to $-\infty$ (Wrightian fitness to $0$), for $r>r^*$ to model the situation where specific binding to the binding site of interest is no stronger than the non-specific mode of attachment:

\begin{equation}\label{Eqn:Fitness}
F(\Delta G(r))=\{
        \begin{array}{c}
          -\frac{1}{2}\kappa_Fr^2\quad\mathrm{for}\ r\le r^*\\
          -\infty\quad\mathrm{for}\ r>r^*\\
        \end{array}
\end{equation}
where $\kappa_F$ is the curvature of the fitness landscape and biologically, roughly corresponds to the strength of selection of this trait; as $\kappa_F$ decreases the fitness landscape becomes more shallow, and so for a fixed effective population size the landscape becomes more neutral. Our choice of fitness function, essentially assumes that for $\Delta G-\Delta G^*>0$, the probability of the TF being bound to its binding site is zero; this is an approximation to the more correct functional form for the proportion of time the TF spends at its TFBS, which will be sigmoidal in form \cite{Gerland2002a} with transition at the critical binding energy $\Delta G^*=\Delta\varepsilon r^*$. However, without a far more detailed model of how occupancy affects gene expression which affects fitness, which is beyond the scope of this work, we are left to choose some arbitrary occupancy threshold below which the organism is inviable; for simplicity, we have chosen $\Delta G^*$, as this threshold naturally corresponds to when specific and non-specific binding are equally like at the site. We expect our qualitative results to be robust to the choice of such a threshold. Similarly, a more detailed consideration would include binding of the TF to other spurious sites in the genome with large sequence similarity; again we expect such consideration will change the value of $\Delta G^*$, but not change the scaling relation $r^*\propto\ell$, as longer binding sites will always have a larger maximum affinity.

To simulate the evolution of TF-TFBS sequence evolution we assume a diploid Wright-Fisher population genetic process with $2N_e$ copies of each gene in the population with a fixed effective population size of $N_e$. As we are in interested in the weak mutation regime ($n\mu_0N_e\ll1$), the simulations consist of a single fixed sequence for the TF-TFBS pair of loci at each time-point, where new mutations either fix or not, decided based on the probability of fixation; this process represents the evolution of a monomorphic population of effective size $N_e$. We assume full linkage disequilibrium within the TF and TFBS loci and linkage equilibrium between loci. In addition, we assume each loci is always homozygous on each lineage; when considering hybrid incompatibilities, the initial product of any cross mating will be heterozygotic at all diverged alleles. All post-zygotic DMIs must be sufficiently deleterious to affect these heterozygotic offspring. There may be many TF-TFBS pairs where the lack of cross binding in heterozygotes does not appreciably change the offspring's viability. We will assume, however, that there are some TF-TFBS pairs that are sufficiently critical such that $r>r^*$ is sufficient to decrease the gene expression level to the extent that the hybrid is inviable; these are the pairs that will be relevant for the speciation process, and therefore are the ones addressed by our model.

We use the Gillespie algorithm \cite{Gillespie1976}, to simulate evolution as a continuous time Markov process; at each step of the simulation the rate of fixation of all one-step mutations from the currently fixed alleles (wildtype) on both TF and TFBS loci are calculated, and one of these mutations is selected randomly in proportion to their relative rate. Time is then progressed by $K^{-1}\ln(u)$, where $K$ is the sum of the rates of all one-step mutants and $u$ is a random number drawn independently between $0$ and $1$, which ensures the times at which substitutions occur is Poisson distributed, as we would be expected for a random substitution process. The rates are based upon the Kimura probability of fixation \cite{Kimura1962}:

\begin{equation}\label{Eqn:Kimura rate}
k=2\mu_0 N_e\frac{1-e^{-2\delta F}}{1-e^{-4N_e\delta F}}
\approx \mu_0 \frac{4N_e\delta F}{1-e^{-4N_e\delta F }},
\end{equation}
where $\delta F$ is the change of fitness of a mutation at a particular location and $2\mu_0N_e$ is the rate at which mutations arise for each amino acid or nucleotide position in a diploid population, where we have assumed that the effective population size is the same as the number of individuals $N$; the latter approximation in Eqn.\ref{Eqn:Kimura rate} assumes $\delta F\ll1$. Note that although in the simulations we use the full form for the fixation probability, typically we would expect fitness effects to be small ($\delta F\ll1$), so the substitution rates only depends on the population-scaled fitness changes $4N_e\delta F$ which, for a given mutation, is proportional to $4N_e\kappa_F$. In the rest of the paper we will refer to the scaled population size $4N_e\kappa_F$ to make it clear that either reducing $N_e$ or $\kappa_F$ (or both) can change the evolutionary outcomes from those dominated by selection to those dominated by drift. For each scaled population size and sequence length, 1000 replicates were run up to a time of $\mu_0t=500$. In addition, simulations were ran up to a shorter time (dependent on the exact value of $4\kappa_FN_e$) with $10^6$ replicates in order to get reliable estimates of the very small probability of a DMI (Fig.\ref{Fig:ProbDMI}) at early times.

\subsection*{A biophysical model of reproductive isolation}

Using the above evolutionary process based on the biophysics of a TF binding DNA, we study allopatric speciation by independently evolving two lineages in the fitness landscape defined by Eqn.\ref{Eqn:Fitness}. We create an ancestral genome containing a protein and a DNA binding site gene, each of length $\ell$, with $\Delta G$ drawn from the equilibrium distribution of binding energies. This ancestral genome is then duplicated, with each copy representing the start of a different isolated population that subsequently evolves independently. If the evolving protein and DNA sequences in one lineage are $\boldsymbol{g}_1^P$ and $\boldsymbol{g}_1^D$ and the other $\boldsymbol{g}_2^P$ and $\boldsymbol{g}_2^D$, we can at each time point calculate the Hamming distance for each hybrid as $h_{12}=d_H(\boldsymbol{g}_1^P,\boldsymbol{g}_2^D)$ and $h_{21}=d_H(\boldsymbol{g}_2^P,\boldsymbol{g}_1^D)$ with corresponding hybrid binding energies, $\Delta G^H_{12}=\Delta\varepsilon h_{12}$ and $\Delta G^H_{21}=\Delta\varepsilon h_{21}$. Using the same fitness function Eqn.\ref{Eqn:Fitness}, we can then evaluate the fitness of the hybrids as a function of time. An incompatibility arises whenever the fitness of the hybrid is $-\infty$ ($h_{12}>r^*(\ell)$ or $h_{21}>r^*(\ell)$), i.e. when a hybrid TF-TFBS specific binding is weak compared the non-specific mode of binding and effectively can no longer recognise its target site. At this point, we assume that the two diverging populations can no longer form viable offspring, and they are reproductively isolated.

This model of TF-TFBS binding energies is inherently epistatic, despite the assumption that the contribution of each pair of interacting pair amino acid and nucleotide is independent and additive to the total binding energy. There is epistasis at both the phenotype and fitness level, the latter due to quadratic selection. Hence, although there is a similarity between our model and polygenic models of quantitative traits under quadratic selection, they are very different as in quantitative traits the phenotype is additive in each loci \cite{Wright1935,Wright1935+,Barton1989}. Here in our model epistasis arises since at each location, say in the protein sequence, whether a given amino acid will give rise to a match or mismatch depends on the particular nucleotide that it is opposite; the binding energy phenotype is a non-linear function of the sequences at the TF and TFBS loci. It is this epistasis that is the source of the Dobzhansky-Muller incompatibilities that we find in our simulations described in the Results section. For example, the common ancestor might be fixed for a pair of sequences $\frac{\A\T\C\G\C}{\A\T\A\G\C}$, which has a binding energy of $\Delta G_{CA}=3k_BT$, as there is only a single mismatch; after a period of divergence, two allopatric populations might be fixed for $\frac{\T\T\A\G\C}{\A\T\A\G\C}$ and $\frac{\A\T\C\G\A}{\A\T\C\G\C}$, each arising from just two substitutions, of compensatory effect, from the common ancestor sequence, so that $\Delta G_1=\Delta G_2=3k_BT$, as there is still only a single mismatch. However, the hybrid sequences are $\frac{\T\T\A\G\C}{\A\T\C\G\C}$ and $\frac{\A\T\C\G\A}{\A\T\A\G\C}$, which correspond to binding energies $\Delta G^H_{12}=\Delta G^H_{21}=6k_BT$, as they each have two mismatches. As the number of substitutions increases on each lineage, we can see that each lineage will maintain good fitness in a stabilizing landscape through compensatory changes, which each try to minimize the number of mismatches; however, each lineage fixes different sets of compensatory mutations, so when combined in a hybrid, the epistasis between pairs of sequences then gives rise to DMIs.

\section*{Results}

\subsection*{Evolution under stabilizing selection on each lineage}

\begin{figure}[ht!]
\begin{center}
{\rotatebox{0}{{\includegraphics[width=0.5\textwidth]{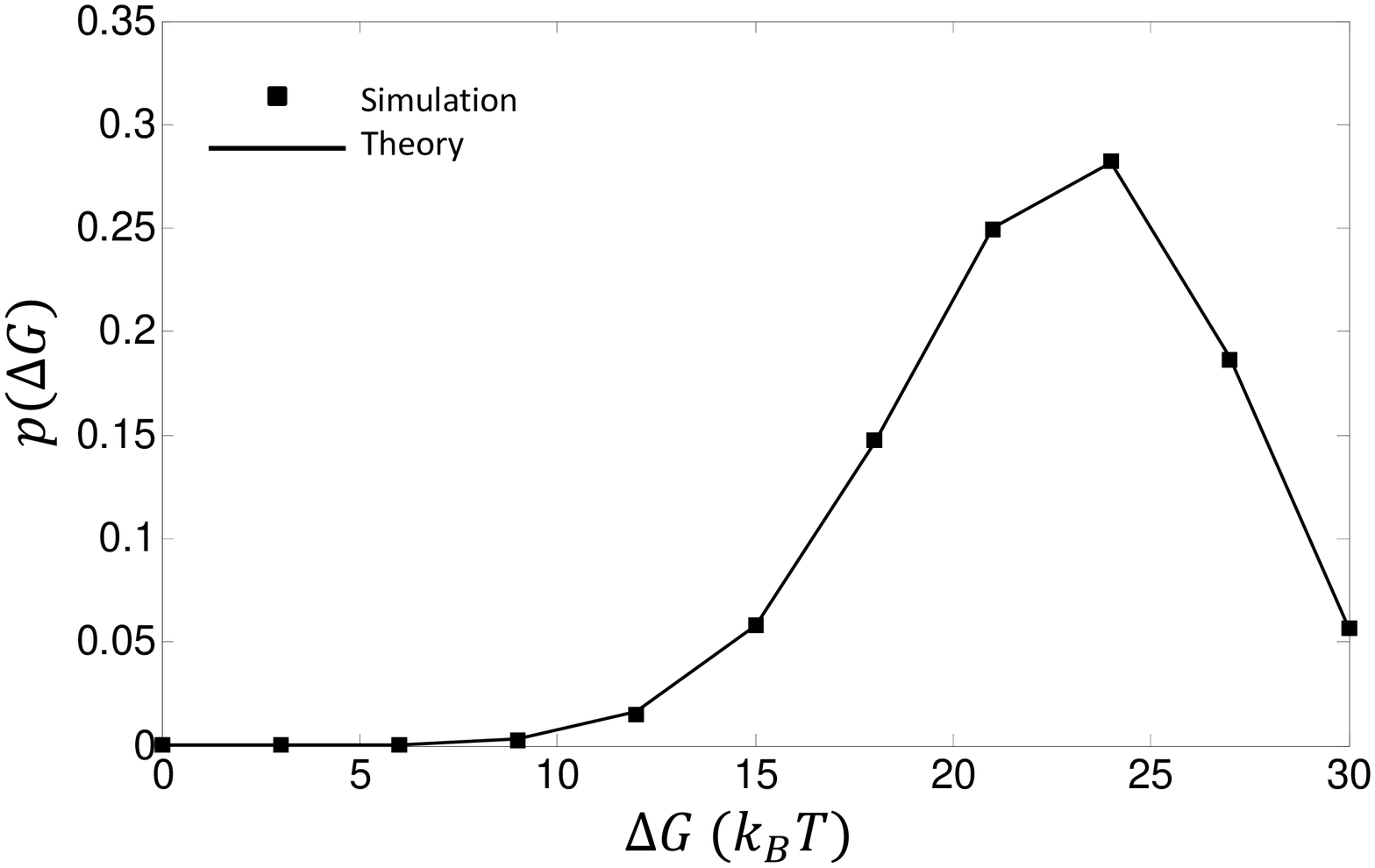}}}}
\caption{Equilibrium neutral distribution of binding energies $\Delta G$ for a 2-state model of TF-DNA binding with a quaternary alphabet for both amino acids and nucleotides. Solid black squares are KMC simulations of neutral evolution, where each sequence is of length $\ell=10$ and the binding energy is $\Delta G=\varepsilon r$, where $r$ is the Hamming distance between the two sequences and $\varepsilon=3k_BT$. We see that under neutral evolution the distribution $\Omega(\Delta G(r))$ is highly non-uniform. The solid line shows the distribution predicted by Eqn.\ref{Eqn:Omega}, which is the relative number of sequences corresponding to a Hamming distance $r = \frac{\Delta G }{\varepsilon}$.
\label{Fig:EntropicDistribution}}
\end{center}
\end{figure}

To understand the qualitative properties of TF-DNA binding evolution, we first consider neutral evolution of such a system and in particular, the resulting distribution of binding energies $\Delta G$. The results of KMC simulations of neutral evolution of sequences $\boldsymbol{g}^D$ and $\boldsymbol{g}^P$ of length $\ell=10$, and $\varepsilon=3k_BT$, with $\kappa_F = 0$ and $r^*=\infty$ (where all sequences have equal fitness) are shown by the solid black squares in Fig.\ref{Fig:EntropicDistribution}. We see that even under neutral evolution of sequences, the distribution of binding energies is non-uniform and roughly Gaussian with a peak between $22 k_BT$ and $23k_BT$. We can understand this by considering the many-to-one mapping often characteristic of the relationship of genotype to phenotype. In particular, there will be a large set of sequences that result in the same binding energy. The number of sequences $\Omega(\Delta G)$ that correspond to a given Hamming distance or energy is non-uniform and given by the binomial distribution

\begin{equation}\label{Eqn:Omega}
\Omega(\Delta G(r))=4^{2\ell}\binom{\ell}{r}\left(\frac{3}{4}\right)^r\left(\frac{1}{4}\right)^{\ell-r},
\end{equation}
where from Eqn.\ref{Eqn:BindingEnergy} $r=\Delta G/\varepsilon$. For example, the number of sequences that give $\Delta G=0$ is $\Omega(\Delta G=0)=4^\ell \approx 10^6$ (for $\ell = 10$), as there is exactly one DNA sequence that matches to each one of the $4^\ell$ protein sequences. This number is very small compared to the number of sequences that have 7 mismatches, $\Omega(\Delta G=21k_BT)\approx3\times10^{11}$, which is close to the mean of the distribution $\langle \Delta G\rangle=3\varepsilon\ell/4=22.5k_BT$. For neutral evolution, the resulting distribution of $\Delta G$ will match the number of sequences corresponding to each $\Delta G$ value. The distribution $\Omega(\Delta G)$ (normalized) expressed in Eqn.\ref{Eqn:Omega} is plotted as a solid line in Fig.\ref{Fig:EntropicDistribution} and we see excellent agreement. This effect of a non-uniform distribution of binding energies of random sequences on evolutionary dynamics has been  well studied \cite{Force1999,Berg2004,Khatri2009,Mustonen2008} and measured empirically for various TFs in \textit{E.coli} and yeast \cite{Mustonen2005,Mustonen2008}; as we will this has a strong impact on the distribution of binding energies under selection at different population sizes.

\begin{figure}[ht!]
\begin{center}
{\rotatebox{0}{{\includegraphics[width=0.5\textwidth]{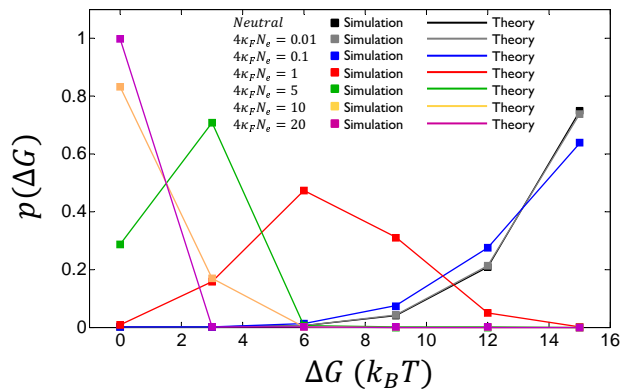}}}}
\caption{Equilibrium distribution of binding energies $\Delta G$ as a result of evolution subject to the quadratic fitness landscape in Eqn.\ref{Eqn:Fitness}, for $\ell=10$; the qualitative results for $\ell=\{5,20\}$ are similar and not shown. The fitness landscape has a fitness cliff (inviability boundary) for $r>r^*=\ell/2=5$ mismatches, or for binding energies greater than $\epsilon r^*=15k_BT$, which represents when the specific binding energy to its binding site is greater than the free energy of binding to the rest of the genome. The solid squares are results of KMC simulations, while the solid lines are the expected distribution from Eqn.\ref{Eqn:Eqm_PDF}, which we see agree very well. In addition, we see that the distribution shifts from one dominated by fitness $F(\Delta G)$ at large population sizes ($4\kappa_FN\gg1$) with a peak at the highest fitness binding energy to one dominated by sequence degeneracy at small population sizes ($4\kappa_FN\ll1$), which is peaked at the inviability boundary, representing the left tail of the neutral distribution in Fig.\ref{Fig:EntropicDistribution} (shown in black).
\label{Fig:EqmDistribution}}
\end{center}
\end{figure}

The distributions of binding energies resulting from evolution with the fitness function Eqn.\ref{Eqn:Fitness} at different scaled population sizes ($4N_e\kappa_F$) are shown in Fig.\ref{Fig:EqmDistribution} for $\ell=10$ and $r^*=\ell/2=5$. The distributions are confined to the region $0\le\Delta G\le \Delta G^*$, where $\Delta G^*=\varepsilon r^*=15k_BT$, is the inviability boundary. Here, it is clear that fitness is not maximized, but instead there is a balance between selection for higher fitness and the tendency to undergo drift towards those phenotypes which correspond to the largest number of sequences. For larger population sizes, selection dominates, resulting in sequence pairs with high fitness. For smaller population sizes, stochastic effects due to drift are more important, resulting in a shift to weaker (more positive) binding energies, approaching the neutral distribution as the population size decreases below the inverse of the overall difference in fitness on the landscape $\frac{1}{2}\kappa_F(\varepsilon r^*)^2$ . This results in a greater effective drift load for smaller population sizes.


The binding energy distributions show that for a general genotype phenotype map fitness is not maximized, but instead there is a balance between selection for higher fitness and the tendency to undergo drift towards those phenotypes which correspond to the largest number of sequences. A powerful approach to dealing with this degeneracy is through the concept of sequence entropy \cite{Barton2009,Khatri2013}, representing the (log) number of sequences encoding a given phenotypic state (e.g. binding energy),

\begin{equation}\label{Eqn:SequenceEntropy}
S(\Delta G)=\ln(\Omega(\Delta G)),
\end{equation}
which is closely related to the Boltzmann entropy from statistical mechanics \cite{Reif}. This entropy measure, should be distinguished from entropies of sequences due to polymorphisms in the population (in this paper we have assumed populations are always monomorphic). The precise combination of fitness and sequence entropy that is maximized during evolution is the function $\Phi(\Delta G)=F(\Delta G)+S(\Delta G)/4N_e$, termed the free fitness \cite{Iwasa1988,Sella2005}, from which the probability density is given by

\begin{equation}\label{Eqn:Eqm_PDF}
p(\Delta G)=\frac{1}{Z}e^{4N_e\Phi(\Delta G)}.
\end{equation}
where $Z$ is a normalization factor, known as the partition function, given by $Z=\sum_{r=0}^\ell e^{4N_e\Phi(\Delta G)}$. This probability density is plotted as solid lines in Fig.\ref{Fig:EqmDistribution} for different population sizes, using Eqns.\ref{Eqn:Omega},\ref{Eqn:Fitness},\ref{Eqn:SequenceEntropy} and \ref{Eqn:Eqm_PDF}; we see that the agreement between the two is excellent.


\begin{figure}[ht!]
\begin{center}
{\rotatebox{0}{{\includegraphics[width=0.5\textwidth]{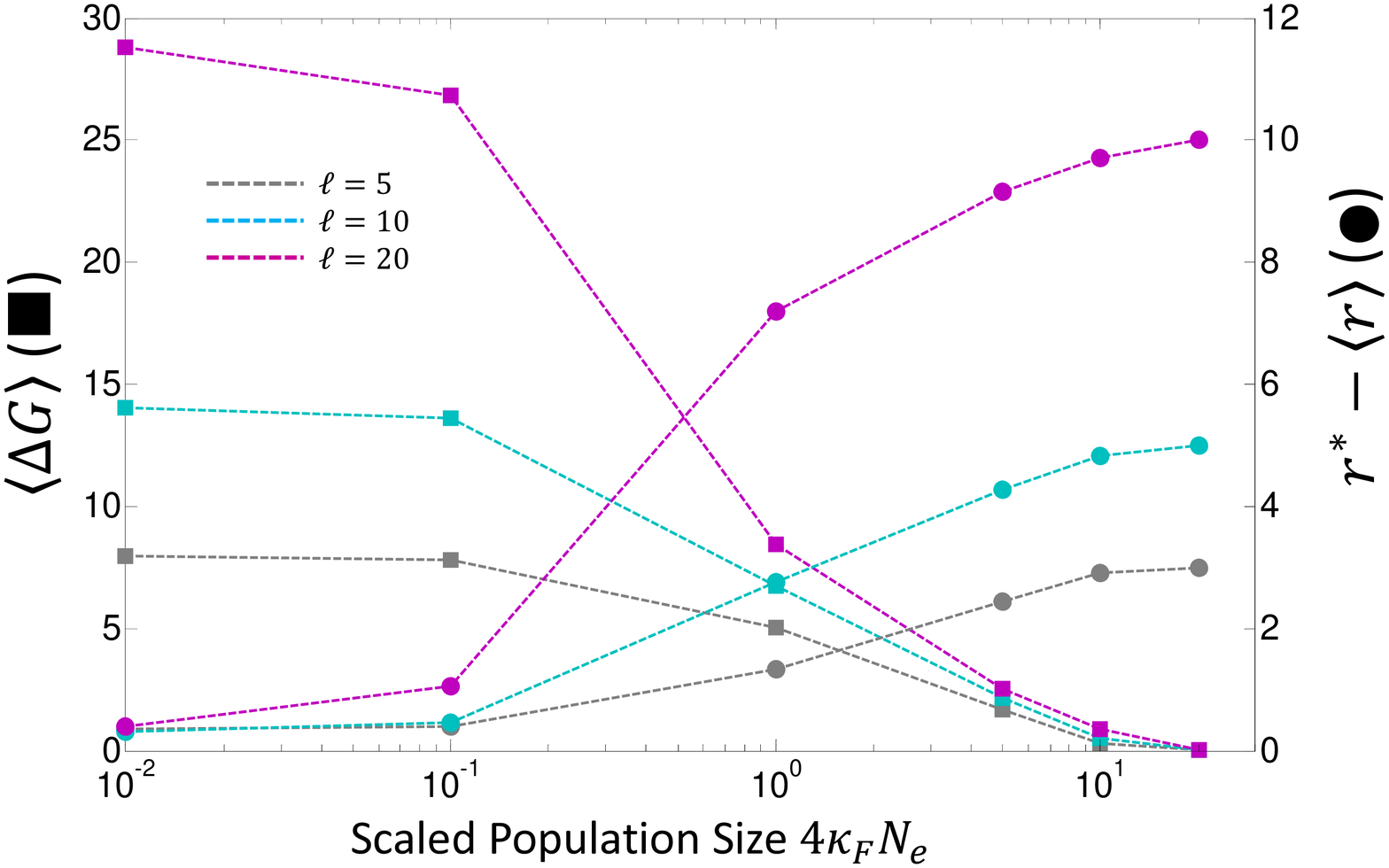}}}}
\caption{Average binding energy, $\langle\Delta G\rangle=\varepsilon \langle r\rangle$, (left axis, squares) and average Hamming distance of populations from inviability boundary, $r^*-\langle r\rangle$, (right axis, circles) as function of scaled population size $4\kappa_FN_e$ and sequence length $\ell$ calculated using KMC simulations. We see that as the population size is decreased the mean hamming distance or binding energy ($\sim$ drift load) increases monotonically and towards the inviability boundary.
\label{Fig:<DG>}}
\end{center}
\end{figure}

This greater drift load is also illustrated in Fig.\ref{Fig:<DG>}, which shows the average binding energy and also the Hamming distance of the populations to the inviability boundary, as a function of the scaled population size $4\kappa_FN$, for sequence lengths $\ell=\{5,10,20\}$; for the corresponding values of $\ell$, we choose $r^*=\{3,5,10\}$, so as to approximately satisfy $r^*=\ell/2$. We see the average binding energy (squares) is larger for smaller population sizes, which corresponds to populations being closer to the inviability boundary as shown by the circles in Fig.\ref{Fig:<DG>}, and hence also a larger drift load. For large population sizes ($4\kappa_FN_e\gg1$), where fitness dominates, the drift load is zero, independent of $N_e$, as $\langle \Delta G\rangle\rightarrow 0$. This means that, as shown in Fig.\ref{Fig:<DG>}, the average Hamming distance to the inviability boundary increases for increasing sequence length --  this arises trivially as $r^*\propto\ell$ -- however, for small population sizes ($4\kappa_FN_e\ll1$) the average Hamming distance to the boundary is roughly independent of sequence length. To understand this we consider that for small populations the distribution is neutral and peaked at the inviability boundary $r^*(\ell)$, as shown in Fig.\ref{Fig:EqmDistribution} and by the fact the mean binding energy is close to $\Delta G^*=\varepsilon r^*$, for $4\kappa_FN\ll 1$ in Fig.\ref{Fig:<DG>}; at the inviability boundary the number of mutations that increase the Hamming distance is just the number of locations that are matched, multiplied by the number of nucleotides that can give a mismatch, $3(\ell-r^*(\ell))=3\ell/2$ and those that decrease it is just the number of mismatched locations, $r^*=\ell/2$. The ratio of these two quantities is independent of $\ell$, showing that there is no net drift bias of the populations at the inviability boundary as $\ell$ changes and so for small populations the average distance to the inviability boundary is roughly independent of $\ell$. As we will see the initial distance of the common ancestor from the inviability boundary has a strong impact on the rate of accumulation of DMIs, as functions of population size and sequence length.

\begin{figure}[ht!]
\begin{center}
{\rotatebox{0}{{\includegraphics[width=0.5\textwidth]{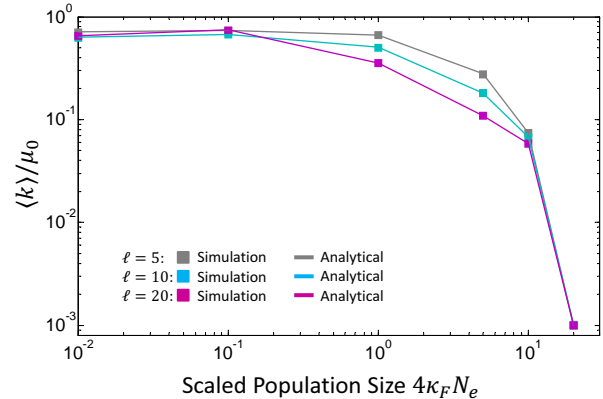}}}}
\caption{Average total substitution rate for both protein and DNA loci, on a single lineage as function of scaled population size $4\kappa_FN$ and sequence length $\ell$. Substitution rate is plotted in units of the nucleotide mutation rate $\mu_{0}$. The solid circles represent KMC simulations, while the solid lines are the theoretical prediction of the average rate $\langle k\rangle=\frac{2N_e\mu_0}{3\ell}\sum_{r=0}^{r^*}p_\ell(r)\left(r\left(\pi^-(r)+\frac{1}{N_e}\right)+3(\ell-r)\pi^+(r)\right)$, where $p_\ell(r)$ is the equilibrium distribution of Hamming distances (shown in Fig.\ref{Fig:EqmDistribution}) and $\pi^-$ and $\pi^+$ are the fixation probabilities for the transition $r\rightarrow r-1$ and $r\rightarrow r+1$, respectively.
\label{Fig:SubstRate}}
\end{center}
\end{figure}

We can also examine the population size dependence of the substitution rate per location in the stabilizing landscape defined by Eqn.\ref{Eqn:Fitness}, as shown in Fig.\ref{Fig:SubstRate} by the solid squares, obtained by simulation for $\ell=\{5,10,20\}$. We see that there is the same qualitative dependence on population size for each sequence length, which can be explained by the average size of fitness effects as the population size changes; at very large population sizes the distribution of binding energies is peaked at $\Delta G=0$ and so the average substitution rate will be dominated by transitions between $r=0$ and $r=1$; forward transitions to $r=1$ happen rarely since the population scaled difference in fitness, $4N_e\delta F=-2\kappa_FN\varepsilon^2$, will be negative with magnitude much greater than 1 for $4\kappa_FN\gg 1$ and so substitutions will occur significantly slower than neutral. While at very small populations although the inverse of the population size is much larger than differences in fitness, since populations spend a large fraction of the time at the inviability boundary $r^*$, the substitution rate is also diminished compared to the expected neutral rate $\mu_0$, since a fraction $(\ell-r^*)/\ell$ of mutations at this boundary are inviable and are never accepted in the population. It is interesting to note that this form of the population size-substitution rate relation is qualitatively similar to what would be expected in a simple stabilizing landscape \cite{Lanfear2014}, however, here at small populations, sequence degeneracy combined with drift pushes populations to the inviability boundary giving rise to an effective substitutional drag relative to the neutral rate.

We find a non-trivial dependence of the substitution rate on sequence length; at large population sizes, as expected, the substitution rate per location is independent of sequence length, but strongly diminished compared to the neutral rate $\mu_0$, as discussed above, due to the discrete changes in fitness being larger than the inverse of the population size. For small populations, we also find that the substitution rate is roughly independent of sequence length; as the distribution of binding energies is peaked at the inviability boundary the substitution rate will be proportional to the number of viable substitutions multiplied by the neutral rate, $\sim\mu_0r^*(\ell)/\ell=\mu_0/2$, which as observed in Fig.\ref{Fig:SubstRate} is independent of $\ell$. However, for intermediate population sizes, where $4\kappa_FN_e\sim1$ the average substitution rate decreases with increasing sequence length. In the large and small populations size limits, all substitutions are either non-neutral or neutral, respectively, for $0\leq r\leq r^*$. However, for intermediate population sizes the quadratic fitness landscape means there is a critical Hamming distance, $r^*_{eff}\approx(4\kappa_FN_e\varepsilon^2)^{-1}$, below which substitutions are effectively neutral ($4N_e|\delta F|\ll1$) and above are non-neutral ($4N_e|\delta F|\gg1$). The effective substitution rate will then be roughly $\sim\alpha(\ell)\mu_0r^*_{eff}/\ell$, where $\alpha(\ell)=\sum_{r=0}^{r^*_{eff}}p_\ell(r)$ is the proportion of time, at equilibrium, spent in the nearly neutral region and $r^*_{eff}/\ell$ is the fraction of nearly neutral substitutions at $r^*_{eff}$; we expect that $\alpha(\ell)$ will decrease for increasing $\ell$, since we find that $p_\ell(r)$ shifts to larger values of $r$ as $\ell$ increases (not shown), due to an increased degeneracy pressure, as the sequence length is increased. So together with the fact that the fraction of nearly neutral mutations decreases for increasing $\ell$, like $r^*_{eff}/\ell$, we see that the average substitution rate is smaller for larger sequence lengths at intermediate population sizes ($4\kappa_FN_e=1$).

\subsection*{Rate of accumulation of hybrid incompatibilities}

\begin{figure}[ht!]
\begin{center}
{\rotatebox{0}{{\includegraphics[width=0.5\textwidth]{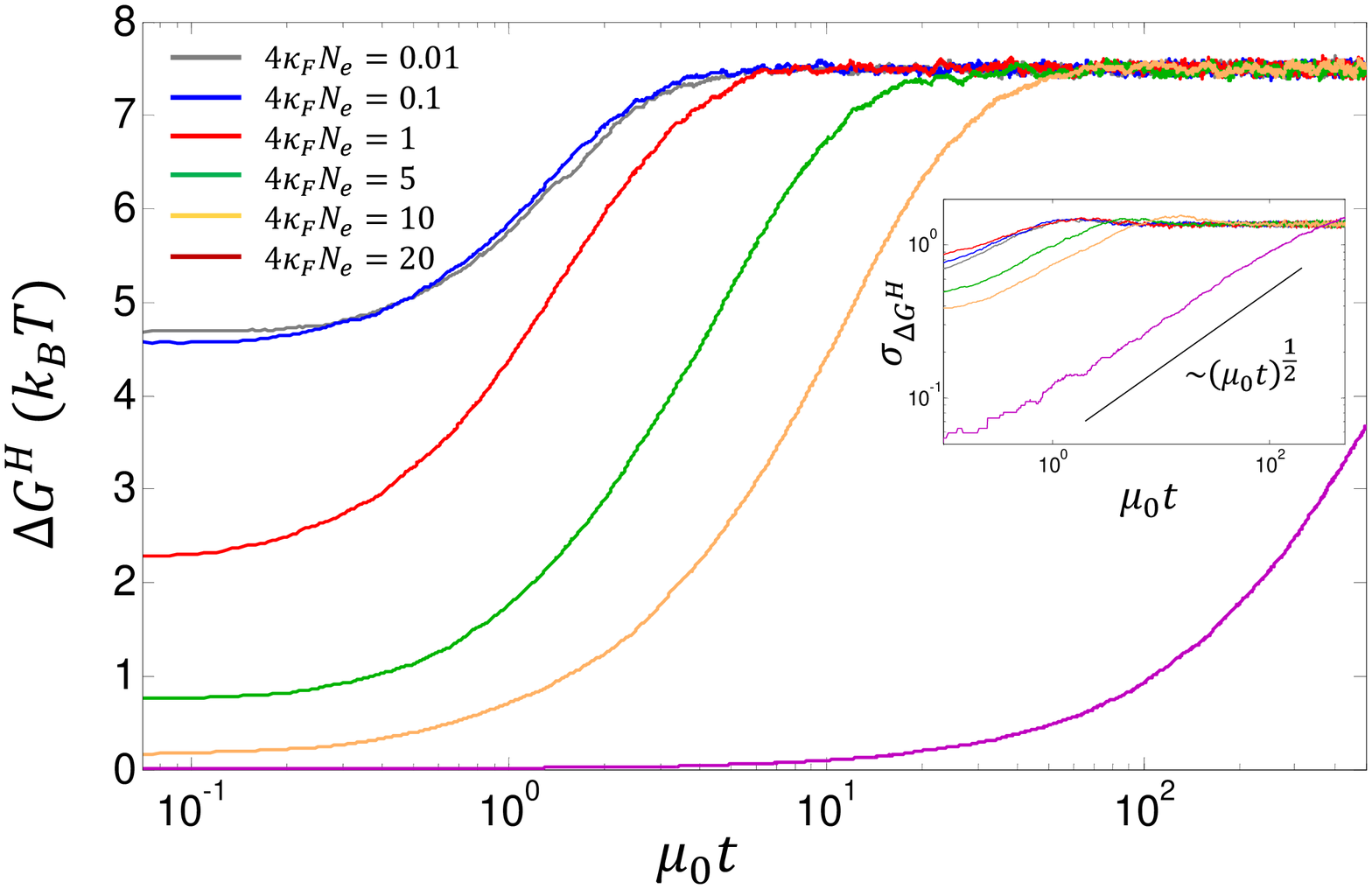}}}}
\caption{Average hybrid binding energy $\langle\Delta G^H\rangle$ as a function of time after divergence from common ancestor $\mu_0 t$ for $\ell=10$ - the qualitative results for $\ell=\{5,20\}$ are similar and not shown. The inset shows the root mean square deviation $\sigma_{\Delta G^H}=\sqrt{\langle(\Delta G^H-\langle\Delta G^H\rangle)^2\rangle}$ of hybrid binding energies as a function of divergence time.
\label{Fig:HybridEnergy}}
\end{center}
\end{figure}

To study the speciation process, we perform replicate simulations of pairs of lineages using the KMC scheme outlined above, with fitness given by Eqn. \ref{Eqn:Fitness}, where each simulation starts with two identical sets of sequences with $\Delta G$ drawn from the equilibrium distribution of binding energies as shown in Fig.\ref{Fig:EqmDistribution}.  We first plot the average hybrid binding energy as a function of $\mu_0 t$ in Fig.\ref{Fig:HybridEnergy}. At zero divergence, the average hybrid binding energies are equal to the average binding energies for that population size, as shown in Fig.\ref{Fig:EqmDistribution}. For long divergence times, the hybrid binding becomes weaker, with the binding energies increasing to a value $\Delta G^H=22.5k_BT$, irrespective of population size, corresponding to the mean of the neutral distribution in Fig.\ref{Fig:EntropicDistribution}; this is exactly what we would expect after a long period of divergence, as protein and DNA sequences from different lineages should have effectively random interactions. The rate at which this neutral distribution is reached depends strongly on population size in an approximately monotonic manner, as would be predicted from the average substitution rate seen in Fig.\ref{Fig:SubstRate}. The inset of Fig.\ref{Fig:HybridEnergy} shows the root mean square, $\sigma_{\Delta G^H}=\sqrt{\langle(\Delta G^H-\langle\Delta G^H\rangle)^2\rangle}$ of hybrid binding energies vs $\mu_0 t$ on a log-log scale; we see that in the limit of large population sizes that $\sigma_{\Delta G^H}\sim\sqrt{\mu_0 t}$, suggesting that the underlying dynamics of the hybrids is effectively diffusive, as suggested by more coarse-grained models \cite{Khatri2013}.

\begin{figure}[ht!]
\begin{center}
{\rotatebox{0}{{\includegraphics[width=0.5\textwidth]{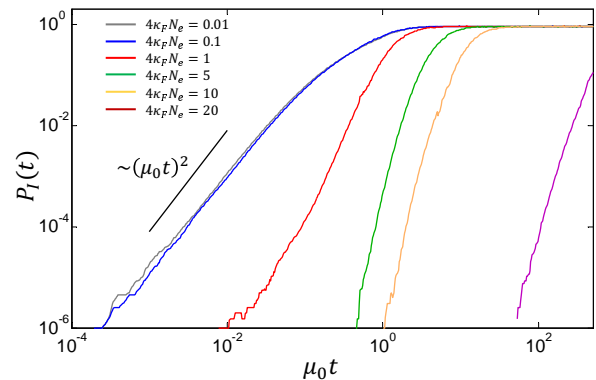}}}}
\caption{Average probability of a DMI as a function of time after divergence from common ancestor $\mu_0 t$ calculated from KMC simulations for various scaled population sizes, for $\ell=10$; the qualitative results for $\ell=\{5,20\}$ are similar and not shown, but the trends with sequence length are demonstrated in Fig.\ref{Fig:TimeToSpeciation}.
\label{Fig:ProbDMI}}
\end{center}
\end{figure}

\begin{figure}[ht!]
\begin{center}
{\rotatebox{0}{{\includegraphics[width=0.5\textwidth]{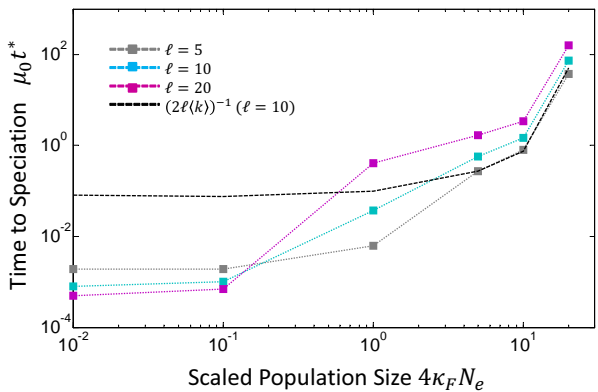}}}}
\caption{Time for reproductive isolation (RI) to arise as a function of scaled population size $4\kappa_FN$ and sequence length $\ell$, defined as the time $t^*$ when the average probability of a DMI crosses a threshold value of $1/M=10^{-5}$, where $M$ is the typical number of interaction partners of a protein in a genome. The black dashed line corresponds to a plot of the inverse of the average substitution rate shown in Fig.\ref{Fig:SubstRate}.
\label{Fig:TimeToSpeciation}}
\end{center}
\end{figure}

The probability of DMIs $P_I(t)$ as a function of $\mu_0 t$ is plotted in Fig.\ref{Fig:ProbDMI}, for various values of $4\kappa_F N_e$ and for $\ell=10$. The qualitative behavior of the plots for $\ell=\{5,20\}$ are similar and for clarity not shown; however, we examine below the dependence of $P_I(t)$ on $\ell$ through the typical time for reproductive isolation to arise in Fig.\ref{Fig:TimeToSpeciation}. We see that the model predicts a very strong population size effect for the dynamics of hybrid incompatibilities; as the population size decreases the timescale for DMIs to arise sharply decreases. This effect saturates for very small population sizes, but diverges for very large population sizes, to the point that reproductive isolation will take extremely long times for very large population sizes ($4N_e\kappa_F\gg10$). This trend can be understood to arise from two effects: 1) as the population size decreases, the initial drift load of the common ancestor is on average larger and so fewer substitutions are required between a pair of divergent lineages for an incompatibility to arise in a hybrid (shown by Figs.\ref{Fig:EqmDistribution}\&\ref{Fig:HybridEnergy}); 2) as the population size increases beyond $4\kappa_FN\sim1$ the substitution rate on each lineage decreases significantly, as seen in Fig.\ref{Fig:SubstRate}, which increases the observed time for incompatibilities to arise. For small population sizes the increase in DMIs is quadratic at small times ($2\ell\mu_0 t\ll1$), while for large population sizes there is a very rapid increase in DMIs, which does not seem to fit a power law and suggests a finite negative curvature on a log-log scale. We note that our prediction at small population sizes and times is the same as Orr's \cite{Orr1995,Orr2001}, but the underlying mechanism in this 2-loci system is very different as it arises from an average over the equilibrium distribution of common ancestor binding energies (Fig.\ref{Fig:EqmDistribution}). The behavior seen at large population sizes is consistent with theoretical predictions of a coarse-grained model of TF-DNA binding evolution \cite{Khatri2013}, where the growth of DMIs is rapid with the asymptotic form, as $t\rightarrow0$ of $P_I(t)\sim \mathrm{erfc}(1/\sqrt{t})\sim \sqrt{t}e^{-1/t}$, which cannot be expressed as a power law for small times. This form arises when considering the distribution of times to diffuse to the incompatibility boundary starting from a fixed binding energy; these are the conditions found for KMC simulations at large population sizes, where hybrid binding energies show neutral diffusive dynamics (inset Fig.\ref{Fig:HybridEnergy}) and the equilibrium distribution for the common ancestor is highly peaked (Fig.\ref{Fig:EqmDistribution}). Finally, we see that at the intermediate population size of $4\kappa_FN_e=1$, there is a transition from the power-law behavior at short times and non-power law at long times, with the transition at approximately $\mu_0t\sim0.1$; this would be as expected if the short-time behavior arises from common ancestors drawn from the right-tail of the probability distribution, near the inviability boundary, for $4\kappa_FN_e=1$ in Fig.\ref{Fig:EqmDistribution}, whilst the long-time behavior arises from common ancestors drawn from around the peak of the distribution, which are further away from the boundary.

In a full genome, where there are many possible interacting genes, it will typically be the short-time behavior of each interacting pair that will dominate. If we assume roughly $m\sim10$ interaction partners per gene and $n_G\approx2\times 10^4$ protein coding genes, we have roughly $M=\frac{1}{2}mn_G\approx10^5$ interaction partners. As only a single one of these interactions giving rise to a DMI is required for RI, we would expect the probability that RI has arisen is $P_{RI}(t)=1-(1-P_I(t))^M$, which at short times is given by $P_{RI}(t)\approx 1-e^{-MP_I(t)}$. In Fig.\ref{Fig:TimeToSpeciation} is plotted the time $t^*$ at which $P_I(t^*)=10^{-5}$, for $\ell=\{5,10,20\}$. We see there is a strong population size dependence on the rate at which RI develops and a weaker, but still significant one on the sequence length. In particular, we see for small populations RI can arise quite quickly, on times where $\mu_0 t^*\approx0.0005$, for $\ell=20$, which corresponds to $\sim250,000$ generations, assuming $\mu_0=2\times10^{-9}$. As discussed above a major determinant at large population sizes on the time for RI to develop is the rate of substitutions on each lineage, the inverse of which is plotted as a dashed line in Fig.\ref{Fig:TimeToSpeciation}; we see that although the inverse substitution rate is a good predictor for large population sizes, for small populations it fails. This is due to the larger drift load for smaller population sizes, which reduces $t^*$ further.

We see that the rate of growth of DMIs and the time for RI to arise has a complicated dependence on the sequence length $\ell$; for small populations sizes ($4\kappa_FN_e\ll1$), RI develops more quickly for longer sequences, whilst for intermediate and large population sizes ($4\kappa_FN_e\ge1$) this trend is reversed and longer sequences mean RI develops more slowly. The divergence rate of the two allopatric populations will be controlled by the total substitution rate for both protein and DNA loci, which is $2\ell\langle k\rangle$; for small populations, this trend arises, trivially, from the fact that the per location substitution rate $\langle k\rangle$ is roughly independent of sequence length (as shown in Fig.\ref{Fig:SubstRate}), giving a higher rate of divergence for larger sequences, together with the fact that the average number of substitutions needed to reach the inviability boundary $r^*(\ell)-\langle r\rangle$ is independent of sequence length (as shown by Fig.\ref{Fig:<DG>}). For large population sizes, RI arises more slowly for larger sequences, despite the fact that, like at small population sizes, the overall divergence rate of the two allopatric populations is larger for longer sequences. If we assume that for large populations the dynamics of the hybrids is diffusive (as suggested by the inset of Fig.\ref{Fig:HybridEnergy}), then the mean square Hamming distance should increase linearly with time $\langle r^2\rangle\sim 2\ell\langle k\rangle t$ \cite{GardinerBook}. We then would expect $t^*\sim \frac{(r^*)^2}{2\ell\langle k\rangle}\sim\frac{\ell}{8\langle k\rangle}$ to increase linearly with $\ell$, as $\langle k\rangle$ is independent of $\ell$ (as shown in Fig.\ref{Fig:SubstRate}); the exact values of speciation times are $t^*=\{37.5,72.1,158\}$, for $\ell=\{5,10,20\}$, so we see that at each doubling of $\ell$, $t^*$ is roughly doubled, lending support to the diffusive model, as well as explaining the trend of a longer $t^*$ for longer sequences in Fig.\ref{Fig:TimeToSpeciation} for large populations. For intermediate populations sizes ($4\kappa_FN_e=1$), we have the same, but stronger trend, which is due to the fact that the per location substitution rate $\langle k\rangle$ is smaller for longer sequences, as shown in Fig.\ref{Fig:SubstRate} and so giving a $t^*$ which grows faster than linear with respect to $\ell$.

\section*{Discussion \& Conclusions}

Dobzhansky, Muller \& Bateson \cite{Dobzhansky1936a,Muller1942,Bateson1909} provided the first solution to Darwin's conundrum of how speciation might arise by suggesting that in allopatry incompatibilities form between co-evolving loci on an epistatic fitness landscape. Many studies have since suggested that the dominant form of reproductive isolation involves the accumulation of Dobzhansky-Muller incompatibilities in geographically isolated populations with no or little gene flow \cite{CoyneOrrSpeciationBook2004,MayrBook1963,Wu1983,Vigneault1986}. The observation of the large diversity of species on small young islands, such as Hawaii \cite{MayrInBook1970}, or on the island of Cuba \cite{Glor2004} and East African Great Lakes \cite{Santos2012,Owen1990}, where in the latter two cases each have been subject to historically fluctuating water levels and thus opportunities for allopatric speciation, suggest that smaller populations speciate more quickly. This is in contrast to lower levels of reproductive isolation observed in marine species with large ranges and population sizes; for example, the relatively small fraction of Pacific-Caribbean species pairs separated by the Isthmus of Panama a few million years ago compared to those that are not reproductively isolated \cite{MayrInBook1970, Mayr1954,Rubinoff1971}. There is also evidence that reproductive isolation arises more slowly in birds compared to mammals \cite{Fitzpatrick2004}. Strikingly, even after roughly $55$Myr divergence \cite{Cooper1997a}, domestic chickens (\textit{Gallus gallus}) can still hybridize with helmeted guineafowl (\textit{Numida meleagris}), where estimates of the effective population size of domestic chickens range between $N_e\approx10^5$ to $10^6$ \cite{Sawai2010}, whereas in contrast, cichlids develop reproductive isolation as quickly as $1-10$Myr after divergence \cite{Stelkens2010} and have relatively small population sizes ($100-10000$ \cite{vanOppen1997,Fiumera2000}). This population size trend is further supported by net rates of diversification \cite{CoyneOrrSpeciationBook2004} inferred from phylogenetic trees \cite{Nee2001,Barraclough2001}. Although, the models of Gavrilets \cite{Gavrilets1999}, Nei \cite{Nei1983} and \cite{Barton1989}, predict that very polygenic traits or those under high mutation rate will tend to show this population size trend, they predict no population size dependence, or are not applicable in the weak mutation regime ($n\mu_0 N\ll 1$). In particular, data on the genetic nature of species differences, suggest many traits involved are oligogenic, involving only a few loci \cite{Orr2001a} and so it is an open question to explain the population size dependence of speciation for such monomorphically evolving traits.

Here, we have developed a biophysically motivated model of how incompatibilities arise in allopatric populations, using a simple model of the co-evolution of transcription factors binding to DNA in the weak mutation, monomorphic regime. A key aspect which this biophysical model of evolution introduces to the picture of fitness landscapes is the idea that many sequences can result in the same phenotype, that is the number of sequences corresponding to each phenotype can be very different, and this uneven distribution can have important consequences for the evolutionary process. As described, our results arise due to a drift-selection balance, which can be cast in the language of a balance between fitness and sequence entropy. The maximum of the free fitness landscape, corresponds to the phenotype when these two evolutionary forces are balanced; importantly, this balance is dependent on the population size. Here, for TF-DNA binding there are many more sequences that have a large number of mismatches compared to those few high fitness sequences that have a small number of mismatches; at smaller population sizes genetic drift dominates pushing the equilibrium towards less fit sequences. This has an important consequence for the dynamics of reproductive isolation, that smaller scaled populations on average have common ancestors with a larger drift load and so a smaller number of substitutions are needed for an incompatibility to arise in hybrids. This leads to the main prediction of the paper that smaller scaled populations ($4\kappa_FN_e\ll1$) develop incompatibilities more quickly. At larger scaled population sizes ($4\kappa_FN_e\gg1$, but still in the weak mutation regime, $n\mu_0 N\ll1$), where fitness dominates drift we find this trend continues, but for a different reason; when $4\kappa_FN_e\gg1$ populations no longer diverge neutrally and instead need to fix deleterious mutants whose difference in fitness is large compared to the inverse of the effective population size. This means that the time for reproductive isolation becomes very long for very large populations. Note however, that although our theory strictly applies to the monomorphic regime, we also expect the effect of sequence degeneracy/entropy to lead to a similar trend of an increasing rate of reproductive isolation for decreasing scaled population size for polymorphic loci, where in addition the effect would be reinforced by the slowed divergence of allopatric lineages due to the mechanism of Gavrilets \cite{Gavrilets1999} and Nei \cite{Nei1983}. In particular, recent work \cite{Tulchinsky2014} with a similar sequence based model, but in the regime where the effect of mutations will be strong, showed that smaller populations are more likely to develop poor hybrid fitness, however, no mechanistic cause is given in their work for this trend and the dynamics of the accumulation of DMIs was not investigated.

We also investigated the effect of sequence length on the rate of developing reproductive isolation. We find that TF-DNA binding with a larger number of nucleotides results in reproductive isolation arising more rapidly for small populations $4\kappa_FN_e\ll1$, but less rapid at intermediate and large populations ($4\kappa_FN_e\ge1$). For small populations, we find the average Hamming distance to the inviability boundary and the average substitution rate are independent of sequence length and so reproductive isolation develops more rapidly because longer binding sites have a larger overall substitution rate and so the two allopatric lineages divergence more quickly. Conversely, when $4\kappa_FN_e\ge1$, despite the same dependence of the average substitution rate on sequence length, longer binding sites are more stable and so require a larger number of mismatches to destabilize the TF to prevent binding to its correct site, we model this simply by having an inviability boundary $r^*\propto\ell$. Guided by our simulation results (inset Fig.\ref{Fig:HybridEnergy}), as well as theoretical studies \cite{Khatri2013}, which suggest that the hybrid binding energies are diffusive, this then suggests that the time for reproductive isolation to arise should grow linearly with sequence length, which we find is in good agreement with our simulations.

Our model then provides a rationale for the observation in the field that smaller populations develop DMIs more quickly, with a robust mechanism that does not require that either lineage pass through a fitness valley. It also, for the first time, provides an insight, through a biophysical model, of the mechanistic causes of how DMIs develop for co-evolved pair-wise molecular interactions. While we would not expect quantitative agreement with biological systems, we can make a rough comparison to empirical data: our results suggest that reproductive isolation can occur on a timescale of order a few hundred thousand generations for small scaled population sizes. Direct studies of interspecific hybrids of African cichlids \cite{Stelkens2010} show that post-zygotic isolation typically arises over a timescale of $\sim4-18$Myr, which corresponds to roughly $\sim1-6$ million generations, assuming a generation time of 3 years \cite{Nagl1998}, which suggests the mechanism we present is consistent with empirical data. Importantly, we see that this mechanism can provide relatively rapid reproductive isolation between lineages with only nearly neutral evolution, without having to invoke positive selection or peak-shifts.

The model studied, however, is simplified compared to the complexity of gene regulation in eukaryotes with multiple TFs binding to enhancers to control gene transcription and each TF having multiple binding sites controlling many different genes. Here, we treat TFs and their binding sites on an equal footing and so for example, the substitution rate in each is the same. It is commonly thought that since TFs are under stronger pleiotropic constraints, they evolve more slowly and so much of the phenotypic divergence between species is driven by cis-regulatory change \cite{King1975a,Wittkopp2008} (and reviewed recently by \citet{Lynch2008a}). We expect that as pleiotropy will act to reduce the substitution rate on a TF, the divergence rate of allopatric lineages will decrease. This suggests that if pleiotropy is important, our simulations may underestimate the average time to reproductive isolation. However, there is increasing evidence that protein evolution driven by protein-protein interaction together, for example, with tissue specific TFs can reduce the pleiotropic constraints on TFs \cite{Lynch2008a}.

Previous theoretical work by Orr \cite{Orr1995,Orr2001} predicts that in the weak mutation regime, the number of incompatibilities should increase as $\sim t^2$ from a fixed common ancestor, due to the combinatorial possibilities over a large number of pair-wise interacting loci. Here, we predict the same growth of DMIs with time, but only for small scaled population sizes ($4\kappa_FN\ll1$) and for a single 2-loci system. However, the underlying mechanism appears to be very different here; the quadratic law arises due to averaging over the distribution of the initial binding energy (or effective drift load) of the common ancestor, which is roughly equivalent to averaging over the growth of DMIs for the different initial drift loads that each pair of loci will have across the whole genome within a single common ancestor. On the other hand, for large populations, which have a peaked distribution of common ancestors relative to the Hamming distance to the inviability threshold $r^*$, we observe that the growth of DMIs does not appear to be described by a simple power law, but instead the results suggest there is a negative curvature to their growth on a log-log plot. In addition, we find that the variance of binding energies increases linearly with time in the limit of large populations (inset Fig.S1), so together with our results that indicate $t^*\sim\ell$, this suggests that from a given common ancestor the hybrid binding energies follow neutral diffusive dynamics. Together, this is as predicted by a simple calculation of the growth of DMIs due to a continuous diffusion model for the evolution of TF-DNA binding \cite{Khatri2013} and arises due to the fact that from a fixed common ancestor there is a finite mutational distance that needs to be diffused by hybrids before incompatibilities can arise; in the low scaled population size limit this behavior turns into a power law when averaged over a broad distribution of common ancestors. We suggest that more detailed studies of species divergence, similar to current works \cite{Matute2010,Moyle2010}, which show a rapid increase in DMIs, should be able to discern between these two qualitatively different behaviors at different population sizes. In particular, recent cross-species ChiP-seq analysis of transcription factor binding \cite{Schmidt2010} suggests a way to explicitly test our predictions at the level of actual binding affinities of hybrid TF-TFBS combinations for recently diverged species, such in the Drosophila family.

The process of speciation underlies the vast diversity of life on Earth. We expect these results to be also seen in more complex models of co-evolving loci since the balance between sequence entropy and fitness poising populations nearer or further away from incompatible regions in a population size dependent manner is likely to be general. Gene expression divergence is thought to underlie many differences between species \cite{King1975,Wolf2010,Wray2007}, for example, in the Galapagos finches \cite{Abzhanov2006}, the various species of \textit{Drosophila} \cite{Wittkopp2008} and with more direct evidence of a role in speciation through the evolution of genes related to transcription factors \cite{Ting1998, Brideau2006}. More recently studies of crosses between \textit{D. melanogaster} and \textit{D. santomea}, which diverged more than 10 million years ago, have revealed how the cryptic divergence of genetic architecture of conserved developmental body plans leads to postzygotic isolation \cite{Gavin-Smyth2013}. Proteins binding to DNA to control gene expression is a prototypical co-evolving system and critical for the proper development of organisms, thus these results have strong implications for speciation rates and diversity of populations at small population sizes. In addition, although our model is motivated by DNA protein binding, the approach could be adapted to any type of interacting macromolecules, for example, co-evolution of protein-protein interactions or the interaction of genes expressed by nucleus and mitochondria, where in particular, such interactions have been shown in yeast to give rise to cytonuclear incompatibilities \cite{Chou2010,Chou2010a}.

\section*{Acknowledgements}
We acknowledge useful discussions with David Pollock, University of Colorado and funding from the Medical Research Council, U.K (funding reference U117573805).

\bibliography{Evolutionpapers}

\end{document}